\documentclass[12pt]{iopart}

\newcommand{\be}{\begin{equation}}
\newcommand{\ee}{\end{equation}}
\newcommand{\bea}{\begin{eqnarray}}
\newcommand{\eea}{\end{eqnarray}}
\newcommand{\bk}{b^{\dagger}}
\newcommand{\ak}{a^{\dagger}}
\newcommand{\ket}[1]{| #1 \rangle}

\newcommand{\av}[1]{\langle #1\rangle}

\newcommand{\pr}[1]{#1^\prime}

\usepackage{psfrag}                  
\usepackage{epsfig}                  

\begin{document}

\title{Effects of interatomic collisions on atom laser outcoupling}
\author{Georgios M Nikolopoulos\dag, P Lambropoulos\dag\ddag and 
	N P Proukakis\S}

\address{\dag\ Institute of Electronic Structure {\rm\&} Laser, FORTH,	
    	P.O.Box 1527, Heraklion GR-71110, Crete, Greece}

\address{\ddag\ Department of Physics, University of Crete,  P.O.Box 2208, 
	Heraklion GR-71110, Crete, Greece}
\address{\S\ Department of Physics, University of Durham, Durham DH1 3LE, United Kingdom}

\begin{abstract}
We present a computational approach to the outcoupling in a simple 
one-dimensional atom laser model, the objective being to circumvent mathematical 
difficulties arising from the breakdown of the Born and Markov approximations. 
The approach relies on the discretization of the continuum representing 
the reservoir of output modes, which allows the treatment of arbitrary forms of 
outcoupling as well as the incorporation of non-linear terms in the 
Hamiltonian, associated with interatomic collisions. 
By considering a single-mode trapped condensate, we study the influence of elastic 
collisions between trapped and free atoms on the quasi steady-state population 
of the trap, as well as the energy distribution and the coherence of the outcoupled 
atoms.
\end{abstract}


\maketitle

\section{INTRODUCTION}
 
The most direct way to create an atom laser, which means an intense directional beam of 
coherent atoms, is based on outcoupling
a pre-formed Bose-Einstein Condensate (BEC). 
Key constituents of such a device are the pumping mechanism replenishing the
trapped condensate as atoms are outcoupled from it and the nature of the
outcoupling mechanism.
This has been demonstrated 
experimentally(for unpumped condensates) leading to the generation of pulsed 
\cite{AtomLaser_Exp1,AtomLaser_Pulsed2,AtomLaser_Pulsed3}
and quasi-continuous 
\cite{AtomLaser_CW1a,AtomLaser_CW1b,AtomLaser_CW2,AtomLaser_CW3,AtomLaser_CW4} 
atom lasers.
There have been numerous theoretical treatments of the output
coupling and atom lasers, and a review of the theory of atom lasers can be 
found in \cite{bal00} and references therein.
Beyond the early rate equation approaches \cite{Rate1,Rate2,Rate3},
these can be crudely classified into mean field approaches and treatments
based on master equations. Initial master equation approaches 
\cite{hol96,wiseman96,guzman,zob98,Wiseman99} were
formulated within the Born-Markov approximation, whereas subsequent work 
 \cite{hopepra97,moypra97,moypra299,hopepra00,jackpra99,bre99} has brought up 
issues pertaining to 
the non-Markovian character of the relevant processes and its differences 
from the optical laser. 
Moreover, not only the Markov but also the Born 
approximation (standard in quantum optics) is not necessarily valid. 
Similar issues have also arisen in recent years in the context of entirely 
different phenomena related to photonic crystals \cite{pl00}.
In particular, the behavior of an excited atom inside such a medium 
exhibiting a photonic band-gap (PBG) around the atomic transition frequency, 
has 
been shown to exhibit a number of features also related to the invalidation 
of Born and Markov approximations for the processes involved. 
As a consequence of the unusual density of states appearing in the 
coupling of the atom to the medium, a master equation which is the standard 
tool in laser theory and quantum optics, can not be derived in that case. 
This rather fundamental limitation has led to the need for alternative 
approaches and techniques, some of which have been proven versatile and useful 
\cite{nikolg99,str99,dio98,gara97,jack01,breMC99,wiseman2002}.

Another common approach to the physics of atom lasers is based on
mean field treatments whereby the dynamics of both trapped and outcoupled 
components
are governed by nonlinear Schrodinger equations, with the two systems coupled
by an external electromagnetic field generating the outcoupling 
\cite{bal96,adams,jap99,jap00,savage2001,savage2002,Graham,Edwards}.
Nonetheless, significant insight regarding the breakdown of Born and Markov
approximations can be obtained by studying the limit of a single-mode
condensate, as demonstrated in 
\cite{hopepra97,moypra97,moypra299,hopepra00,jackpra99,bre99}.
Although in the limit of no interactions between the trapped and the outcoupled atoms, 
the corresponding master equations can be solved exactly, it is still not clear, 
despite recent work \cite{jackpra99}, what is
the effect of interactions between trapped and outcoupled atoms.
In the present work, we show that one 
particular approach 
\cite{nikolg99,nikolg00,nikolg01,nikolg02}, which originated in the context of 
atomic decay in a PBG medium, can be useful in addressing this and other
related questions   
pertaining to the outcoupling in atom lasers. 
The approach we have in mind relies on the discretization of continua 
appearing in the equations of motion that couple the few degrees of freedom, 
usually referred to as ``the system'', to a reservoir which, by its nature, 
contains infinitely many degrees of freedom corresponding to a part of the 
Hamiltonian with a continuous spectrum. 

In the following sections, we apply the discretization technique in the 
context of a one-dimensional model for an atom laser, with the limitation 
to one dimension due only to computational constraints. The
current paper does not aim at describing the physical processes occurring 
in the recently-observed effectively one-dimensional systems 
\cite{MIT_1D,Schreck,Ott,Hansel,Schneider,lowD} which are
prone to large phase fluctuations \cite{Gora,Stoof},
an issue that is dealt with in \cite{Stochastic}.
Our findings are therefore relevant to and of interest for current regimes
of three-dimensional atom laser operation 
\cite{AtomLaser_Exp1,AtomLaser_Pulsed2,AtomLaser_Pulsed3,AtomLaser_CW1a,AtomLaser_CW1b,AtomLaser_CW2,AtomLaser_CW3,AtomLaser_CW4}. 
We first investigate the validity of the method and study the dynamics of the 
trapped as well as the outcoupled atoms, in the absence of interatomic collisions 
between trapped and outcoupled atoms. 
We show that, in that context, the method allows us to treat practically exactly the 
problem, even for a range of parameters where it can be treated approximately.
Then, we extend the treatment beyond the majority of master equation 
approaches that have 
been addressed so far in the context of atom lasers, by allowing interatomic 
collisions between the trapped and the untrapped (free) atoms. We discuss the 
assumptions under which such processes can be incorporated in our simulations 
and study the system dynamics under their influence.
     
\section{OUTCOUPLING FREE OF INTERATOMIC COLLISIONS}
\label{secII}
We consider a BEC consisting of a large number of bosonic atoms cooled into a 
single eigenmode of a trap. 
The atoms are coherently coupled out of the trap by the application of 
external electromagnetic fields which induce an atomic transition from 
the internal state $(\ket{t})$ of the trapped atoms to an 
untrapped state $\ket{f}$ \cite{bal96,adams,Graham,Edwards,ban99}.
Initially, only the lowest mode of the trap is populated 
(condensate mode). We may further ignore the population of higher modes 
even at later times, if the frequency and the linewidth of the applied fields 
are chosen appropriately \cite{moypra97}. 

The dynamics of the system are described by a Hamiltonian of the form
\be
H=H_S+H_R+V,
\label{hamilt}
\ee
where $H_S$ and $H_R$ correspond to the trapped (system) and 
untrapped (reservoir) atoms, respectively, while $V$ refers to the output 
coupling.

Introducing a set of bosonic operators $\{\ak, a\}$ for the representation 
of the condensate, the corresponding Hamiltonian reads
\be
H_S=\hbar\omega_0\ak a,
\label{hamilt1}
\ee
where $\omega_0$ is the condensate-mode frequency.
Accordingly, the part of the Hamiltonian for the free atoms is 
\be
H_R=\hbar\int_{-\infty}^{+\infty}dk \omega_k\bk_k b_k,
\label{hamilt2}
\ee
where $\bk_k$, $b_k$ are the corresponding bosonic operators, with $k$ being 
the momentum of the center-of-mass motion. 
 
\subsection{Output coupling}
\label{ssecA}
The condensate is coupled to the reservoir through the interaction term 
\be
V=\hbar\int_{-\infty}^{+\infty}dk g(k)(\bk_k a+b_k\ak).
\label{hamilt4}
\ee
In general, the form of the coupling constant $g(k)$ is determined by the 
type of mechanism applied to coherently couple the trapped atoms out of the 
trap (output coupler) \cite{Graham}. 
For a Raman output coupling mechanism, the outcoupled state mimics the
trapped condensate at short times \cite{adams,Edwards}. Assuming
 Gaussian like profile for the 
ground state of the harmonic trap\cite{hopepra97,moypra97}, the coupling 
constant can be approximated by
\be
g(k)=\frac{\sqrt{\Gamma}}{(2\pi\sigma^2)^{1/4}}\exp(-k^2/4\sigma^2),
\label{coup}
\ee
where $\Gamma$ and $\sigma$ denote the strength and the width of the coupling, 
respectively.  
Additionally, the frequency $\omega_k$ is related to the corresponding 
momentum by $\omega_k=\hbar k^2/2m$. 
It is this quadratic dependence of $\omega_k$ on $k$, which together with 
the functional dependence of $g(k)$ on $k$, give rise to mathematical 
difficulties even in this simple model. Note that, the dependence 
of $\omega_k$ on $k$ is in fact identical to the analogous dispersion 
relation in PBG materials, which is responsible for the 
mathematical difficulties in that context. 

For the one-dimensional problem under consideration, the density of states which are 
available to a free atom can be determined by the dispersion relation as follows
\be
\rho(\omega)=\bigg|\frac{dk}{d\omega}\bigg|=
	\sqrt{\frac{m}{2\hbar\omega}}\Theta(\omega),
\label{dos}
\ee
where $m$ is the atomic mass and $\Theta(\omega)$ the usual step function.
Taking advantage of the symmetrical shape of the coupling and the even 
parity of $\omega_k$, we may reduce the $k$-space only to the $k>0$ sub-space. 
The spectral response of the output coupling is then of the form
\be
{\cal D}_G(\omega)=2|g(\omega)|^2\rho(\omega)=\frac{C}{\pi}\frac{\exp[-
	\omega/\alpha]}{\sqrt{\omega}}\Theta(\omega),
\label{srG}
\ee
where the effective coupling constant $C$ is given by 
\be
C=\Gamma\sqrt{\frac{\pi}{\alpha}}\qquad \textrm{and} 
	\qquad \alpha=\frac{\hbar\sigma^2}{m}.
\label{CC}
\ee
As $\sigma$ and $\Gamma$ tend to infinity, we obtain the broad-band limit 
of equation (\ref{srG}),
\be
{\cal D}_{BB}(\omega)=\frac{C}{\pi}\frac{1}{\sqrt{\omega}}\Theta(\omega),
\label{srBB}
\ee 
which is identical to the corresponding spectral response of PBG continua.
Note that our model does not consider any losses due to gravity or 
Raman-momentum kick which could, however, have been included.

Given the Hamiltonian (\ref{hamilt}) one may proceed to derive the 
Heisenberg equations of motion for the operators of interest, examples 
of which are discussed below. At this point, let us recall that in problems 
where the Born and Markov approximations are valid, as for example an 
excited atom decaying spontaneously in the vacuum of open space, the 
reservoir (continuum) variables can be eliminated. This procedure leads to a 
Markovian master equation governing the evolution of the system. This is what 
can not always be done in the present context. In the following subsection 
we present, for the model Hamiltonian under consideration, an approach capable 
of providing not only the time dependence of the number of atoms in the 
condensate, but also the distribution of the outcoupled atoms in frequency 
domain, irrespective of the strength of the coupling between the system and 
the reservoir. Moreover, it is applicable to more general coupling shapes 
and suitable for the inclusion of additional interaction terms in the 
Hamiltonian (see section \ref{secIII}) for a more realistic description of the 
physical situation.   

\subsection{Discretization of the reservoir}
\label{ssecB}
In order to deal with the structured reservoir, we follow the discretization 
approach developed in the context of PBG continua 
\cite{nikolg99,nikolg00,nikolg01,nikolg02}. 
Briefly, we substitute the reservoir for frequencies within a range 
around $\omega_0$ $(\omega_{low}<\omega<\omega_{up})$, by a number $(N)$ 
of discrete modes, 
while the rest of the atom-mode density is treated perturbatively since 
it is far from resonance. In general, there is no unique way of discretizing 
a continuum. For the sake of illustration, in the present work we apply two 
different discretization schemes for the Gaussian-like profile and its 
broad-band limit, respectively.

Specifically, for the Gaussian-like coupling we adopt a uniform discretization 
scheme, choosing the frequencies of the modes to be 
\be
\omega_j=\omega_{low}+j\delta\omega,
\label{disc1a}
\ee   
where the mode spacing $\delta\omega$, is determined by the upper-limit 
condition of the discretization, namely
\be
\omega_{up}=\omega_{low}+N\delta\omega.
\label{disc1b}
\ee
The corresponding coupling for the $j$ mode, is determined by the 
spectral response (\ref{srG}) as follows
\be
{\cal G}_j^2={\cal D}_G(\omega_j)\delta\omega.
\label{disc1c}
\ee

Alternatively, for the broad-band coupling (equation \ref{srBB}), we may choose the 
integral form of equation (\ref{disc1c}), obtaining thus frequency-independent 
couplings for the modes, determined by
\be
{\cal G}^2N=\int_{\omega_{low}}^{\omega_{up}}d\omega{\cal D}_{BB}(\omega),
\label{disc2a}
\ee
while the frequency for the $j$ reservoir atom-mode is given by 
\be
\omega_j=\omega_{low}+j^2\delta\omega,
\label{disc2b}
\ee
with the spacing $\delta\omega$ determined by 
\be
\omega_{up}=\omega_{low}+N^2\delta\omega.
\label{disc2c}
\ee
Finally, note that in $k$-space the discretization would rely on the 
following equations 
\be
k_j=k_{low}+j\delta k,\quad {\cal G}_j=g(k_j)\delta k,\quad k_{up}=k_{low}+j\delta k,
\ee
from which we may obtain equations (\ref{disc1a})-(\ref{disc2c}), using the 
definition of the density of atomic states (equation \ref{dos}).

\subsection{Heisenberg equations of motion}
\label{ssecC}
In the Heisenberg picture, the evolution of an arbitrary operator ${\cal A}$ 
is governed by 
\be
\frac{d{\cal A}}{dt}=-\frac{i}{\hbar}[{\cal A}, H],
\label{hem}
\ee
where $H$ is the Hamiltonian, which for the problem at hand is 
determined by equations (\ref{hamilt})-(\ref{hamilt4}).
Thus, for the operators of the system and the reservoir we obtain the following 
equations of motion
\bea
\frac{da}{dt}&=&-i\omega_0a-2i\int_0^\infty dk g(k)b_k dk,
\label{em1a}\\
\frac{db_k}{dt}&=&-i\omega_k b_k-i g(k)a,
\label{em1b}\\
\frac{d(\ak a)}{dt}&=&-2i\int_0^\infty dk g(k)(\ak b_k-\bk_k a).
\label{em1c}
\eea

The solution of the above differential equations can be obtained in terms 
of the inverse Laplace transform. Furthermore, as has been noted in 
\cite{hopepra97,moypra97,moypra299}, the solution for the mean number of atoms in 
the condensate 
$(\av{\ak(t) a(t)})$ can be written as $\av{\ak(t) a(t)} = \av{\ak(t)}\av{a(t)}$,  
where $\av{a(t)}$ can be obtained as solution of equation (\ref{em1a}), assuming the 
initial condition $\av{a(0)}=\sqrt{\av{\ak(0) a(0)}}$, while 
$\av{\ak(t)}=\av{a(t)}^*$. 
This assumption stems from the fact that the initial state of our model 
corresponds to a BEC in the atomic trap, which is a coherent state 
$\ket{\beta}$ of definite global phase and thus 
$\av{\ak(0) a(0)}=|\beta(0)|^2=\av{a(0)}^*\av{a(0)}$.
It is the bilinear form of the Hamiltonian, however, that preserves this coherence in 
time. 
What has to be noted here is that, irrespective of the initial conditions, 
we may obtain the evolution of $\av{\ak(t) a(t)}$ by 
combining equation (\ref{em1c}) with the following equations of motion for 
$\ak(t) b_k(t)$ and $\bk_k(t) b_q(t)$ respectively:
\bea
\frac{d(\ak b_k)}{dt}&=&-i(\omega_k-\omega_0)\ak b_k-ig_k\ak a+
	2i\int_0^\infty dq g(q)\bk_q b_k,\\
\label{em2b}
\frac{d(\bk_k b_q)}{dt}&=&-i(\omega_q-\omega_k)\bk_k b_q + 
	i g(k)\ak b_q - i g(q)a \bk_k.
\label{em2c}
\eea

Discretizing the continuum over a range of frequencies 
$(\omega_{low}<\omega<\omega_{up})$, the integrals are converted to sums over 
discrete 
modes of frequencies $\omega_{j(m)}$, with $j(m)$ running from $1$ to $N$, 
while the number of discrete modes covers only part of the reservoir. 
Assuming, as must be the case, that the discretized range is sufficiently 
large to incorporate the essential effects of the system-reservoir coupling, 
the remaining part of the continuum 
$(\omega<\omega_{low}\quad\textrm{and}\quad\omega>\omega_{up})$ can 
be eliminated adiabatically to order of $|g(k)|^2$, in the standard 
fashion. 
Specifically, the derivative of $b_k$ in equation (\ref{em1b}) for $\omega_k$ 
outside the discretized range, is set equal to zero, yielding the approximate 
solution
\be
b_k(t)\approx-\frac{g(k)}{\omega_k}a(t),\quad \textrm{for} \quad \omega_k<\omega_{low}\quad\textrm{and}\quad\omega_k>\omega_{up}. 
\label{elimBk}
\ee 
These approximate solutions are then fed back into the equations for the 
operators in the discretized part of the continuum. 
As a consequence of this adiabatic elimination, a shift appears in the 
equations involving derivatives of the operators $a(\ak)$ and 
$\ak b_k(a\bk_k)$. Furthermore, since the equations of motion 
(\ref{em1c}) - (\ref{em2c}) constitute a closed set of differential 
equations 
we may deal with expectation values of the operators belonging to the 
discretized space of the continuum, obtaining  
\bea
\frac{d\av{\ak a}}{dt}&=&2\sum_{j=1}^N {\cal G}_j\Im\{\av{\ak b_j}\},\\
\label{em3a}
\frac{d\av{\ak b_j}}{dt}&=&-i(\omega_j-\omega_0+S)\av{\ak b_j}-
	i{\cal G}_j\av{\ak a}+
	i\sum_{m=1}^N {\cal G}_m\av{\bk_m b_j},\\
\label{em3b}
\frac{d\av{\bk_m b_j}}{dt}&=&-i(\omega_j-\omega_m)\av{\bk_m b_j} + 
	i {\cal G}_m\av{\ak b_j} - i {\cal G}_j\av{\ak b_m}^*,
\label{em3c}
\eea
where for simplicity of notation, $\omega_r$, ${\cal G}_r$ and $b_r$ stand 
for $\omega_{k_r}$, ${\cal G}_{k_r}$ and $b_{k_r}$ respectively, 
with $r\in\{j,m\}$, while $\Im\{\cdot\}$ denotes the imaginary part of the 
expression inside the curly braces. 
The shift term is given by 
\be
S=\int_{0}^{\omega_{low}}d\omega\frac{{\cal D}(\omega)}{\omega}+	
	\int_{\omega_{up}}^{\infty}d\omega\frac{{\cal D}(\omega)}{\omega},
\label{shift}
\ee
where ${\cal D}(\omega)={\cal D}_{G(BB)}(\omega)$ and the lower bound of the first 
integral has been set to zero as 
dictated by the step function in the spectral responses of the reservoir. 
Furthermore, in the above set of equations we have used the fact that 
for each pair of arbitrary operators ${\cal A}$ and ${\cal B}$ we have 
$({\cal A}{\cal B})^\dagger={\cal B}^\dagger{\cal A}^\dagger$ and thus 
$\av{{\cal A}{\cal B}}^*=\av{{\cal B}^\dagger{\cal A}^\dagger}$. 
The same equality allows us to deal with half of the matrix elements 
$\av{\bk_m b_j}$, those which involve $m\leq j$ since the elements involving 
$j<m$ can be obtained by $\av{\bk_m b_j}=\av{\bk_j b_m}^*$.

In the following sections, we present some selected results for the above 
discretized set of differential equations. Specifically, for various values of the 
parameters and for both Gaussian and broad-band couplings, we investigate the time 
dependence of the number of atoms in the condensate and the energy 
distribution of the outcoupled atoms $\av{\bk_k b_k}$. 

At this point, some general remarks need to be made concerning the expected 
accuracy of the discretization approach. 
Two main parameters enter the system of equations, namely the 
number of modes $N$ and the upper(lower) limit of the discretized 
part of the continuum.
The number of modes will determine the maximum time for which the numerical 
solution of the discretized system can be considered accurate. This is already 
known from the experience gained in PBG media. If one wants to obtain a 
correct solution for longer times (and this is the case for larger 
values of coupling, when a steady state is reached more slowly), one has to 
increase the number of modes and the number of differential equations, 
which scales roughly as $N^2$. 

\begin{figure}
  \begin{center}
    \leavevmode
    \epsfxsize8.5cm
    \epsfbox{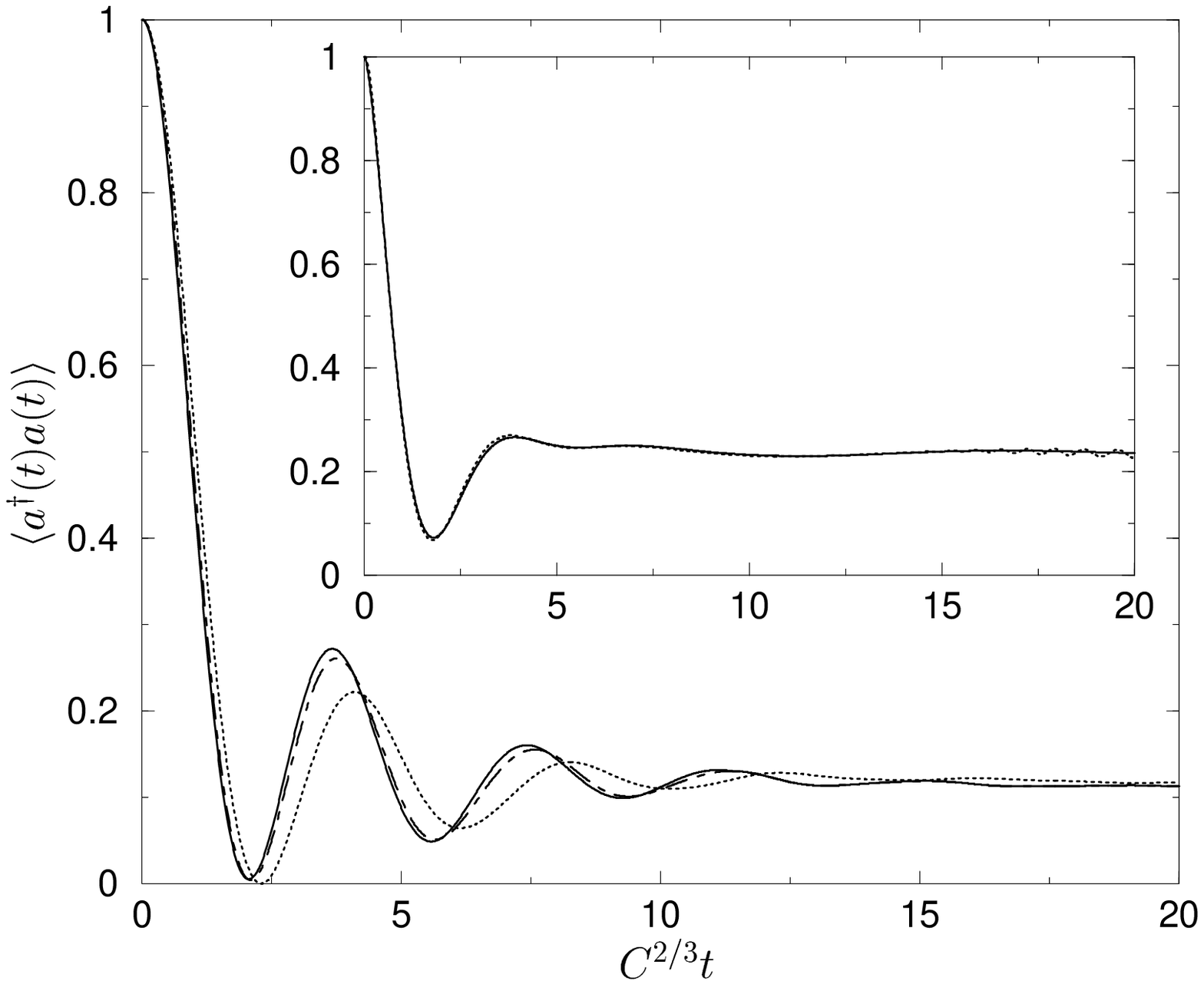}
  \end{center}
\caption{
	The normalized mean number of atoms in the condensate as a function of the 
	dimensionless time $C^{2/3}t$ for Gaussian $(\Gamma=10^6 sec^{-2})$ 
	and broad-band (inset) couplings. The solid lines correspond to the 
	exact solutions obtained by means of the Laplace transform. We also 
	plot the discretization solutions 
	for $N=50$, $\omega_{low}=0$ and $\omega_{up}=3.0C^{2/3}$ (dotted line) 
	as well as for $N=1000$, $\omega_{low}=0$ and $\omega_{up}=3.0C^{2/3}$ 
	(dot-dashed line), respectively. 
	Inset: $N=50$, $\omega_{low}=0$ and $\omega_{up}=10.0C^{2/3}$ (dotted line)
}
\label{convergence.fig}
\end{figure}

On the other hand, the parameters $\omega_{up}$ and $\omega_{low}$ must be chosen 
so that $\omega_{up}-\omega_{low}$ is large enough to include all modes with 
significant role (for sure it must be chosen such that 
$\omega_{low}<\omega_0<\omega_{up}$) but also such that the 
spacing $\delta\omega$ will produce a smooth enough energy spectrum.
As a general remark pertaining to the results discussed below, 
we note that the convergence of our calculations has been checked in terms of 
the number of discrete modes as well as the range of discretization. 
As is depicted in figure \ref{convergence.fig}, in order to reproduce the results 
obtained by means of the Laplace transform in the context of Gaussian 
couplings(solid line) \cite{moypra299}, we find that we need at least $500$ discrete 
modes (dot-dashed line) and a value of $3$ for $\omega_{up}$ in units of $C^{2/3}$. 
As far as the short-time and the long-time behavior are concerned 
(in which we are mainly interested), we may obtain the dynamics of the system even 
for $50$ or $60$ modes and $\omega_{up}=3$ (dotted line), avoiding thus 
time-consuming numerical calculations which are expected to add quantitative  
and not qualitative corrections to our results. 
For the broad-band coupling (inset) the situation is slightly different, in the 
sense that the solution obtained with $50$ discrete modes (dotted line) is almost 
indistinguishable from the exact solution (solid line). 
Nevertheless, as a consequence of the discretized continuum, 
it exhibits revivals for longer times, which are expected to appear for later and 
later times as we increase the number of discrete modes. 

To facilitate comparison with the results of Moy et al. \cite{moypra97,moypra299}, 
throughout this paper we consider $m=5\times 10^{-26}kg$, 
$\omega_0=2\pi\times 123sec^{-1}$ and $\sigma=10^6 m^{-1}$.  
Since, however, the time scale on which the main dynamics take place is 
expected to be strongly dependent on the coupling parameters, 
from now on we present the dynamics of the system as a function of the dimensionless 
time $C^{2/3}t$.  

\subsection{Dynamics}
\label{ssecD}
\begin{figure}
  \begin{center}
    \leavevmode
    \epsfxsize8.5cm
    \epsfbox{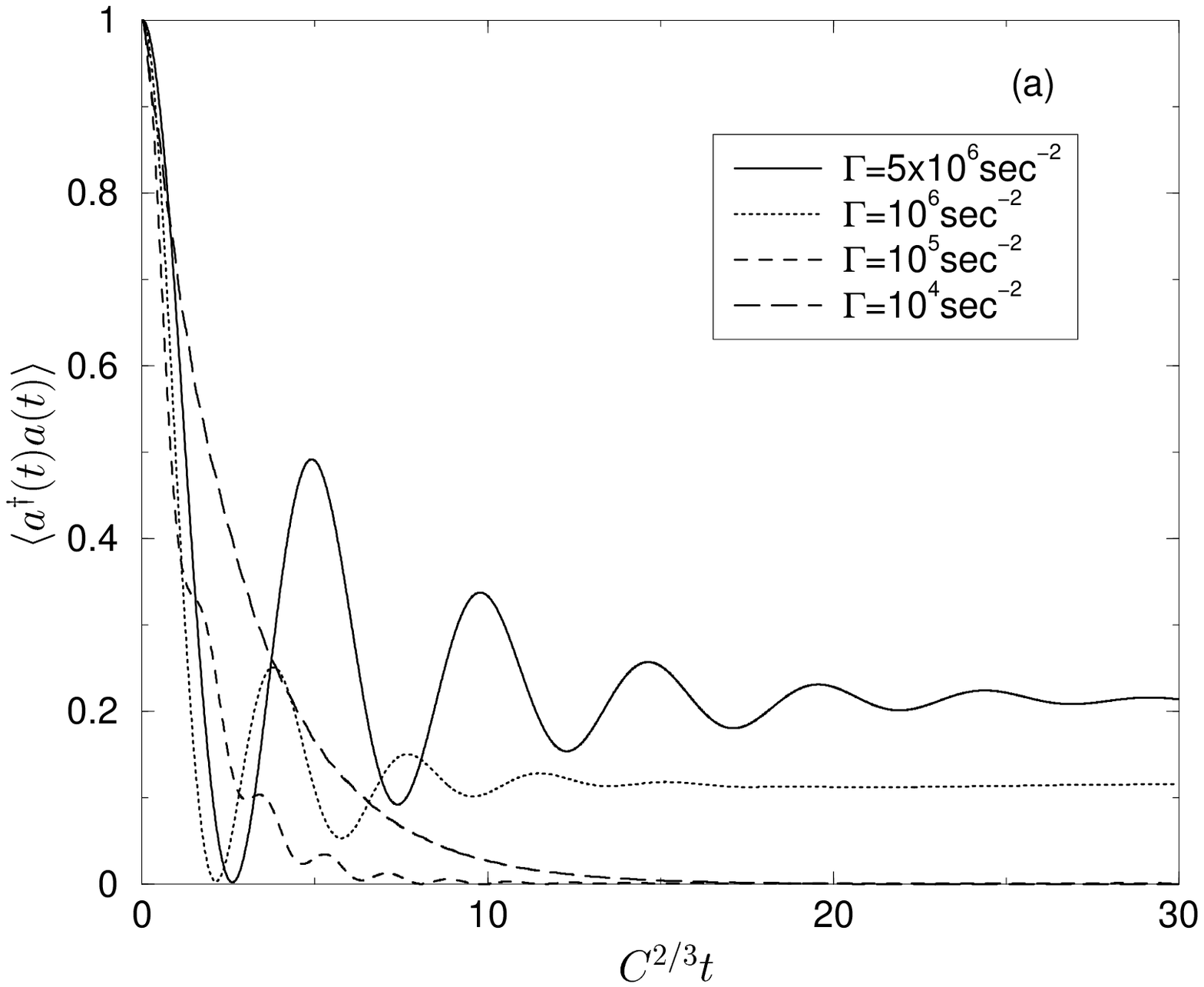}
  \end{center}
  \begin{center}
    \leavevmode
    \epsfxsize8.5cm
    \epsfbox{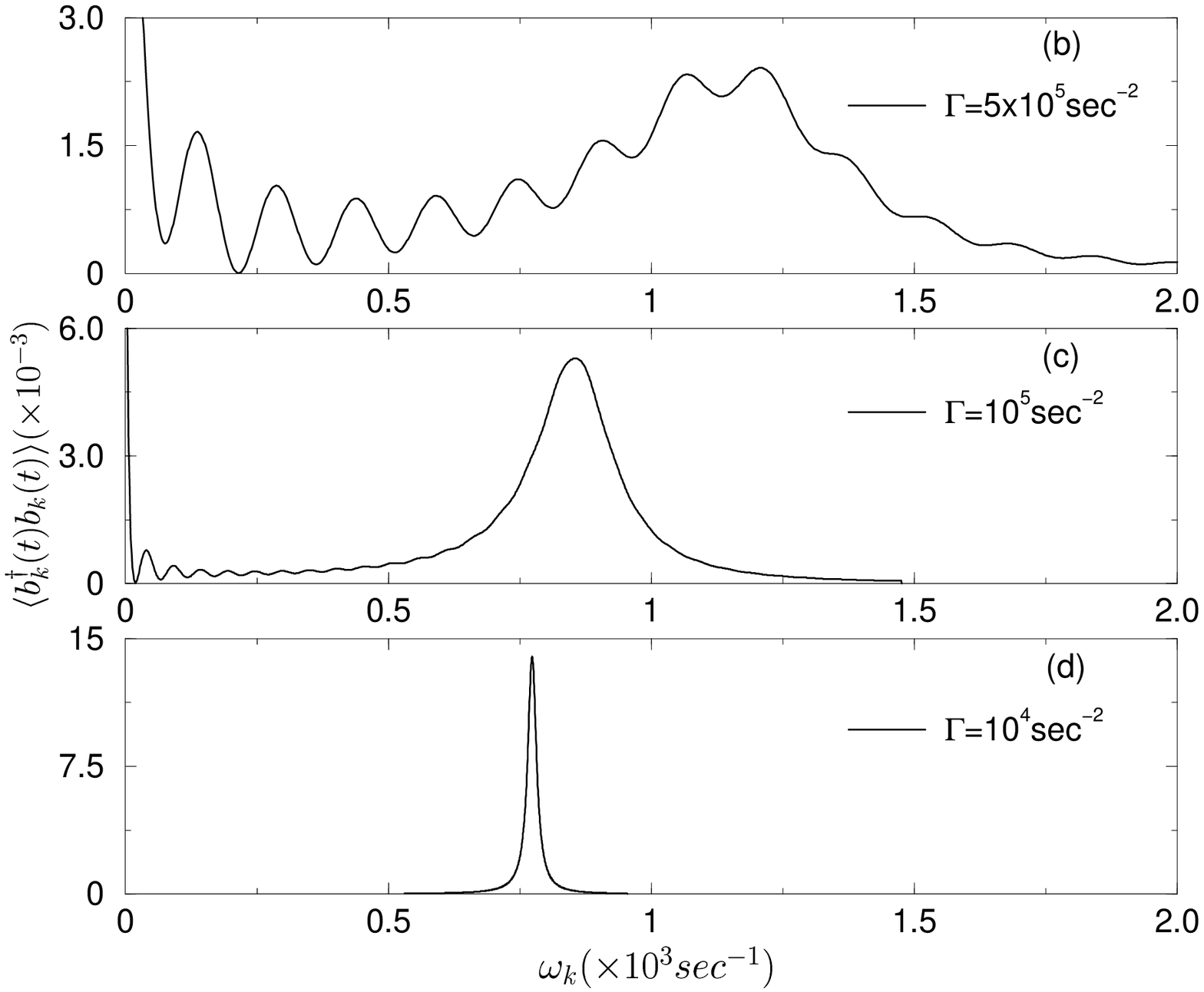}
  \end{center}
\caption{The evolution of the normalized trap population, as a function of the 
	dimensionless time $C^{2/3}t$ (a) and the energy distribution of 
	the outcoupled atoms  at $t=30C^{-2/3}$ [(b)-(d)], for Gaussian coupling 
	and various  coupling strengths.
	The calculations are for 1000 discrete modes, $\omega_{low}=0$ and 
	$\omega_{up}=18C^{2/3}$ $(\Gamma=10^4sec^{-2})$, 
	$\omega_{up}=6C^{2/3}$ $(\Gamma=10^5sec^{-2})$, 
	 $\omega_{up}=3C^{2/3}$ $(\Gamma=5\times 10^5-5\times 10^6 sec^{-2})$.
}
\label{variusGammaG.fig}
\end{figure}

In figure \ref{variusGammaG.fig}(a), we present the time evolution of the normalized 
mean number of atoms (population) in 
the condensate mode for Gaussian coupling and various values of $\Gamma$. 
As was expected, for weak couplings in relation to $\omega _0$, the trap mode 
exhibits a Markovian behavior in the sense that its population decays exponentially 
to zero(long-dashed line). 
As we increase the coupling strength, however, the trap population begins 
exhibiting non-Markovian features. 
Specifically, after a transient regime where part of the 
initial population is lost, the system undergoes Rabi-like oscillations. 
These oscillations can be interpreted as a beating of the system between two 
different paths associated with the two characteristic frequencies 
of the problem, namely the trap-mode frequency $\omega_0$ and the singularity 
$\omega=0$, where the spectral response is peaked. 
The crucial role of these two characteristic points in the system dynamics, is also 
reflected in the energy distribution of the outcoupled atoms 
[figures \ref{variusGammaG.fig}(b)-(d)] which, for relatively 
large couplings, exhibits peaks at $\omega=\omega_0$ and $\omega\approx 0$, 
respectively.  In the weak-coupling regime, the peak at $\omega_0$ is prominent. 
As we enter the strong coupling regime, however, a larger part of the continuum  
is involved and thus, on the one hand the distribution becomes broader 
while on the other, the Gaussian coupling tends to 
couple the cavity mode not only to modes with frequencies in the vicinity of 
$\omega_0$, but also to low-energy modes $(\omega\approx 0)$. 
Furthermore, the central frequency of the distribution is shifted towards higher 
frequencies. This behavior is analogous to the shift in the levels of a 
two-level atom coupled to a photonic continuum and in the context of 
atom lasers has also been noted by Jeffers and co-workers \cite{jefpra00}.   

\begin{figure}
  \begin{center}
    \leavevmode
    \epsfxsize8.5cm
    \epsfbox{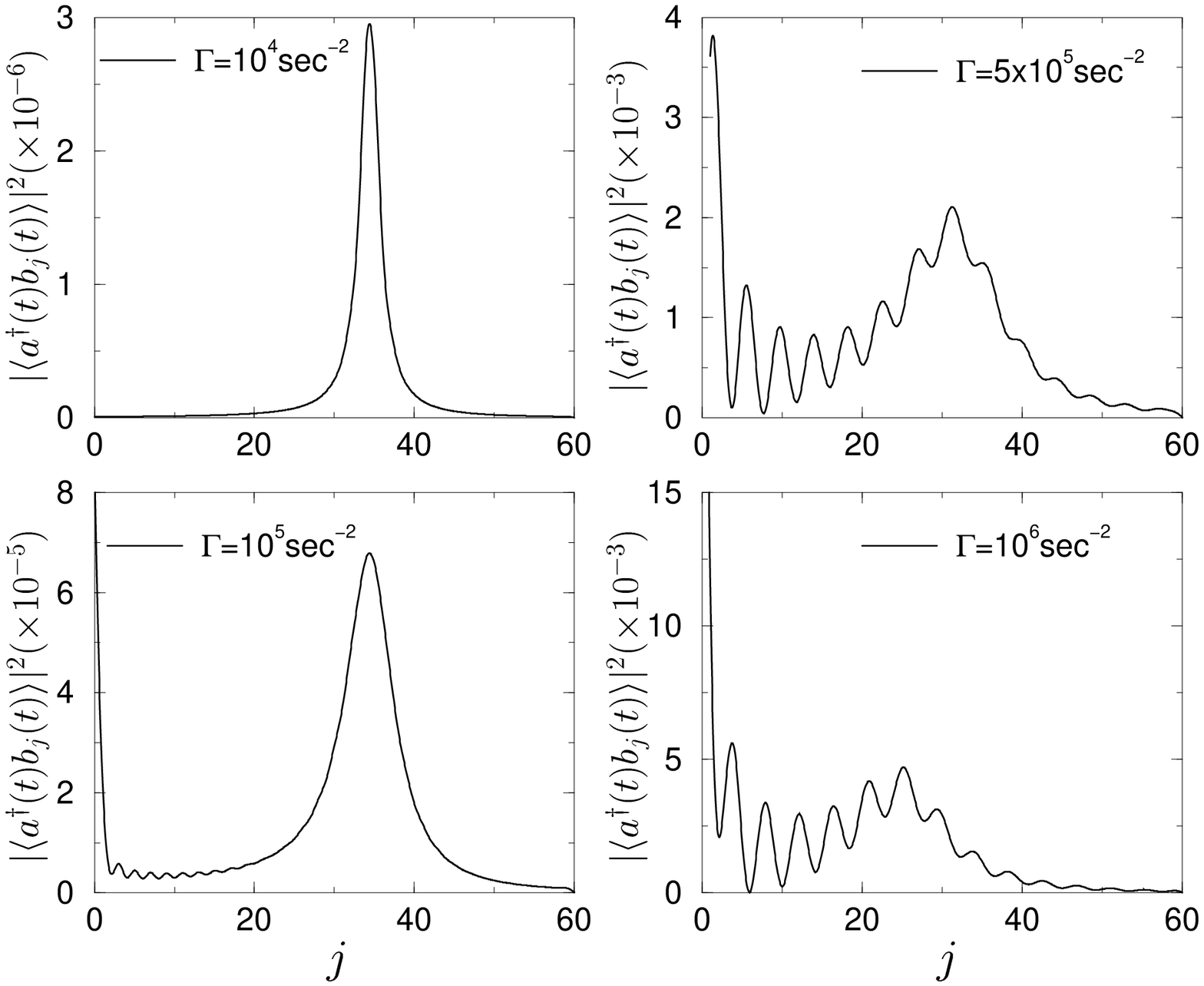}
  \end{center}
  \begin{center}
    \leavevmode
    \epsfxsize8.5cm
    \epsfbox{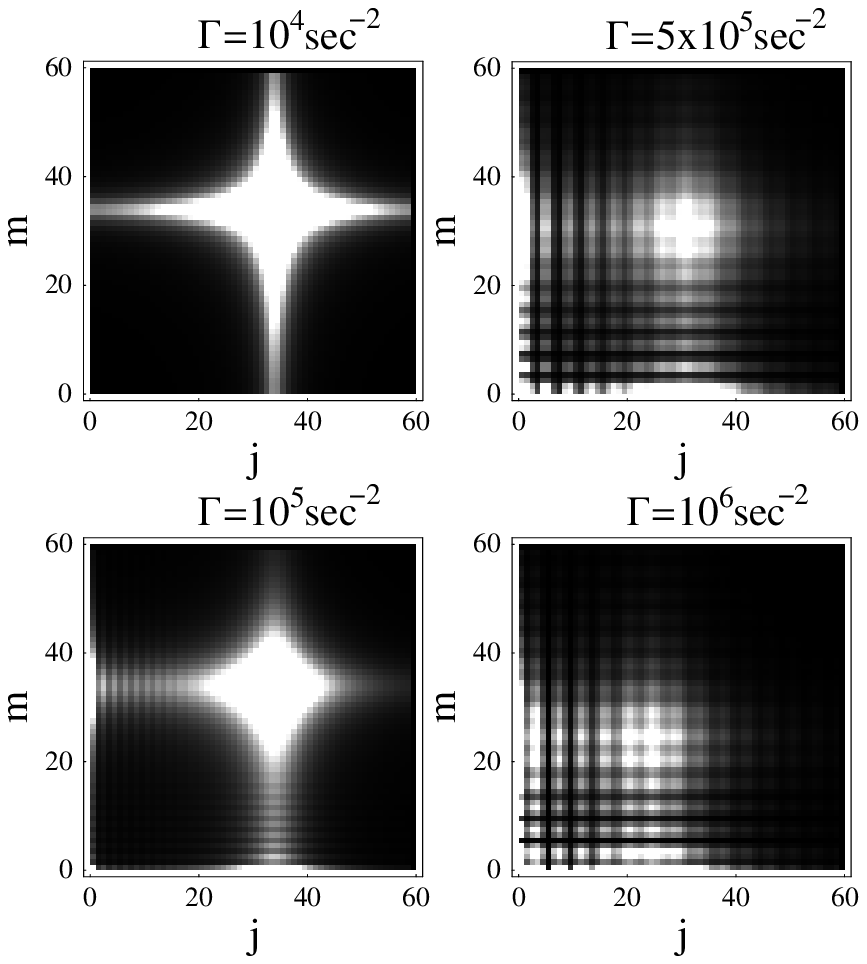}
  \end{center}
\caption{The magnitude of the matrix elements $\av{\ak(t)b_j(t)}$ (upper) and 
	$\av{\bk_m(t)b_j(t)}$ (lower) for Gaussian coupling at $C^{2/3}t=30$  
	and various coupling strengths $\Gamma$. 
	For the lower graphs, dark and white regions represent 
	matrix elements with negligible and large magnitude respectively.  
	The calculations are for $60$ discrete modes and cutoff frequencies:
	$\omega_{low}=10C^{2/3}, \omega_{up}=18C^{2/3}$ $(\Gamma=10^4sec^{-2})$; 
	$\omega_{low}=0, \omega_{up}=6C^{2/3}$ $(\Gamma=10^5sec^{-2})$; 
	$\omega_{low}=0, \omega_{up}=3C^{2/3}$ 
	$(\Gamma=5\times 10^5-10^6 sec^{-2})$. 
}
\label{coherenceG.fig}
\end{figure}

From the physical point of view, the oscillations in the trap population stem from 
the outcoupled atoms which are fed back into the trapped state, reexciting thus the 
condensate mode. 
Due to this back action of the reservoir, the system finally reaches a steady state 
where the condensate mode is partly excited. 
With increasing coupling strength, the oscillations become more pronounced 
and faster, while the population trapping in the cavity increases.
As is evident in figure \ref{coherenceG.fig}, the non-zero steady-state trap 
population is also associated with strong coherence between the atoms in the trap 
and the outcoupled atoms (upper graphs), as well as between the free atoms 
(lower graphs). 
Although for weak couplings the coherence is negligible and is restricted 
to the vicinity of the trap-mode frequency, it increases rapidly for 
increasing values of the coupling strength, while simultaneously it spreads out to 
a larger number of modes. This behavior clearly indicates the departure of the 
system from the Markovian dynamics. 
In figure \ref{coherenceG.fig} we note that the index of the discrete mode 
corresponding to the trap-mode frequency $\omega_0$, is given by 
\be
j_0=\Bigg[\frac{(\omega_0-\omega_{low})N}{\omega_{up}-\omega_{low}}\Bigg],
\label{j0}
\ee 
where $[x]$ denotes the integer part of $x$.

\begin{figure}
  \begin{center}
    \leavevmode
    \epsfxsize8.5cm
    \epsfbox{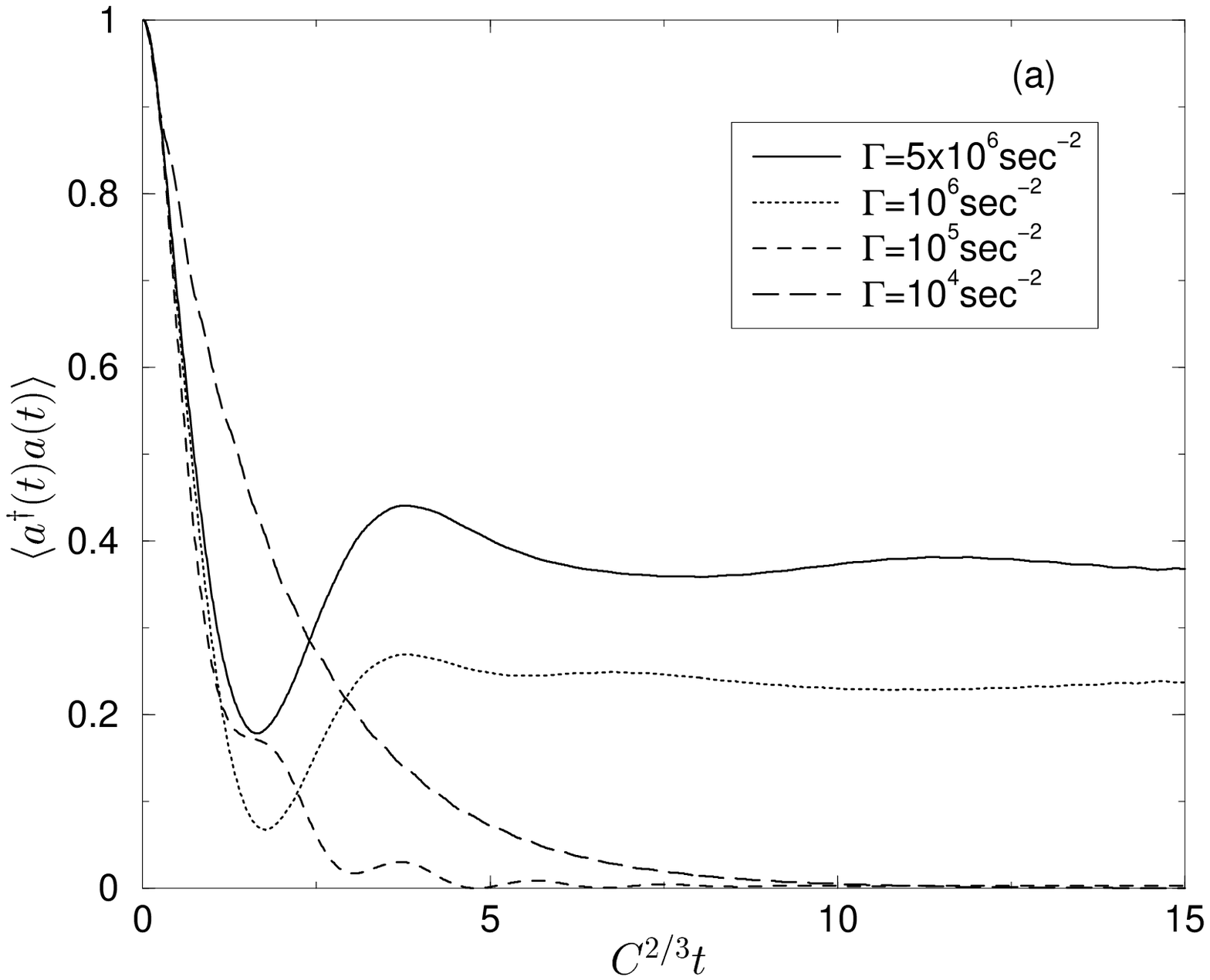}
  \end{center}
  \begin{center}
    \leavevmode
    \epsfxsize8.5cm
    \epsfbox{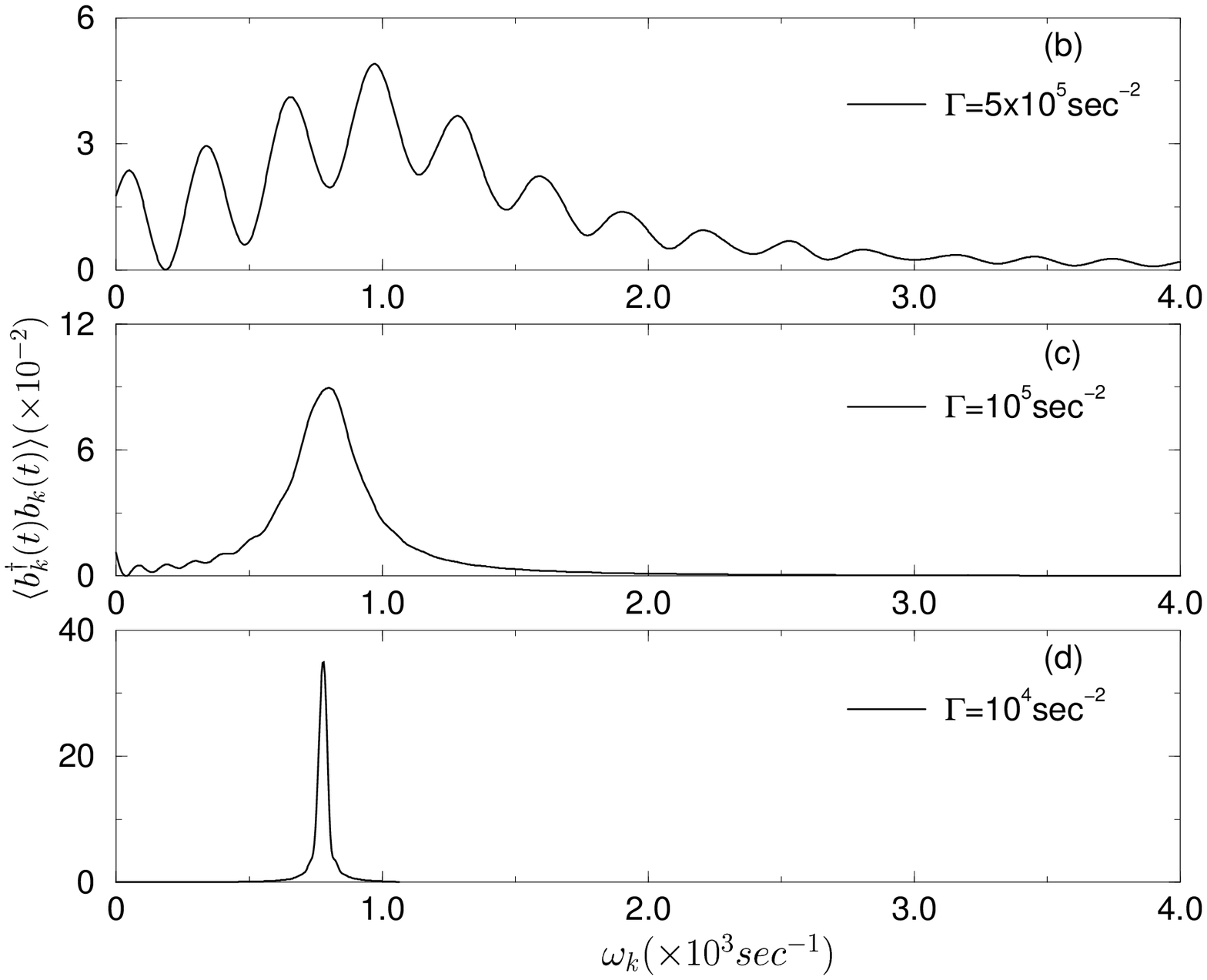}
  \end{center}
\caption{The evolution of the nomalized trap population, as a function of the 
	dimensionless time $C^{2/3}t$ (upper graph) and the energy distribution of 
	the outcoupled atoms  at $t=15C^{-2/3}$ (lower graph), for broad-band 
	coupling and various  coupling strengths.
	The calculations are for $100$ discrete modes, $\omega_{low}=0$ and 
	$\omega_{up}=20C^{2/3}$.
}
\label{variusGammaBB.fig}
\end{figure}

For the broad-band coupling (figure \ref{variusGammaBB.fig}), the dynamics are 
basically  the same. 
The  broad-band nature of the 
coupling, however, allows a significant interaction with more continuum states. 
Thus, the system reaches steady state much faster, while the 
oscillations in the cavity population are not so pronounced. 
Moreover, the energy distribution of the outcoupled atoms is much broader than in 
the Gaussian coupling.  
  
\section{INTERATOMIC COLLISIONS}
\label{secIII}
Up to now we have not considered any collisional interactions between trapped and 
free atoms in our model. As a result, we had to deal with a Hamiltonian bilinear 
in the field operators and thus with a 
closed set of linear differential equations for the operators of interest. 
In real condensates, however, interatomic collisions take place and affect, 
in most cases, the dynamics of the system significantly. 
Specifically, considering a small rate of output from the trap, the output atomic 
beam is dilute and thus, we may neglect collisional 
interactions between free atoms. We may further ignore inelastic collisions between 
untrapped and trapped atoms, since the corresponding 
mean free path is much larger than the dimensions of a typical atomic condensate in 
the trap. 

In our subsequent discussion, we focus on the elastic collisions of the untrapped 
atoms with those in the trap. Considering low-energy $s$-wave 
scattering of length $a_s$ between trapped and free atoms, we may adopt a standard 
hard-core interatomic potential which for a three-dimensional model in position 
representation is 
\be
{\tilde U}_{tf}({\bf r}-{\bf r}^\prime)=U_{tf}^{(3D)}\delta({\bf r}-{\bf r}^\prime),
\label{hamilt3.1a}
\ee
where $U_{tf}^{(3D)}=4\pi\hbar^2 a_s / m$.
For the model at hand, however, we have to derive an one-dimensional, approximate 
form of equation (\ref{hamilt3.1a}), capable of providing us with an estimate of the 
importance of interactions in our model. 
In the case of tight transverse confinement, the scattering strength can be
well approximated by integrating out the transverse directions leading to 
\cite{Olshanii}
\be
{\tilde U}_{tf}(x-x^\prime)\approx\frac{U_{tf}^{(3D)}}{2\pi a_{ho}^2}\delta(x-x^\prime)=2\hbar a_s\omega_{\perp}\delta(x-x^\prime),
\label{hamilt3.1}
\ee
where $\omega_{\perp}$ the transverse confining frequency and $a_{ho}=\sqrt{\hbar/m\omega_{\perp}}$ 
the corresponding harmonic-oscillator length. 
In this paper we are not concerned with the limit of reduced dimensionality \cite{MIT_1D}
and we have only limited ourselves to one-dimensional simulations for computational ease.
The corresponding part of the Hamiltonian then takes the form
\be
V_{tf} =\hbar\int_{-\infty}^{+\infty}dk\int_{-\infty}^{+\infty}dq 
	{\cal \kappa}(k,q)\bk_k b_q \ak a,	
\label{hamilt3.2}
\ee 
where, for a Gaussian ground state of the trap, the coupling 
constant ${\cal \kappa}(k,q)$ reads,
\be
{\cal \kappa}(k,q)={\cal N}a_s\omega_{\perp}\exp\biggr[-\frac{(k-q)^2}{8\sigma^2}\biggr],
\label{hamilt3.3}
\ee
with ${\cal N}$ being the number of trapped atoms. Note that the coupling constant 
is symmetric with respect $k$ and $q$ and thus $V_{tf}$ is Hermitian.
  
In analogy to the output coupling, the spectral response for interparticle interactions 
can be defined as
\bea
\fl {\cal M}_{G}(\omega,\omega^\prime)&=&4|{\cal \kappa}(\omega,\omega^\prime)|^2\rho(\omega)
	\rho(\omega^\prime)=C_M\frac{\exp[-(\sqrt{\omega}-
	\sqrt{\pr{\omega}})^2/2\alpha]}{\sqrt{\omega\pr{\omega}}}
	\Theta(\omega)\Theta(\omega^\prime),
\label{AA}
\eea
where the effective interatomic coupling $C_M$ is 
\be
C_M=\frac{2{\cal N}^2 a_s^2\omega_{\perp}^2 m}{\hbar}.
\label{CM}
\ee
As $\sigma$ and $\Gamma$ tend to infinity, we obtain the broad-band limit 
of equation (\ref{AA}): 
\be
{\cal M}_{BB}(\omega,\omega^\prime)=
	C_M\frac{1}{\sqrt{\omega\pr{\omega}}}\Theta(\omega)\Theta(\omega^\prime).
\label{AAbb}
\ee
For $a_s\approx 1 nm$ and $m\approx 5\times 10^{-26} kg$, 
we have $C_M\approx 10^{-9}{({\cal N}\omega_\perp)}^2 sec$.
A reasonable estimate of realistic interatomic coupling strengths in our model, 
can be obtained by noting that, for ${\cal N}\approx10^3$ atoms and 
$\omega_{\perp}\approx 2\pi\times 1000 sec^{-1}$, a typical value for $C_M$ is 
$C_M\approx10^4 sec^{-1}$, or else $C_M\approx 10 C^{2/3}$. 

\subsection{Heisenberg equations of motion}  
Having now included interatomic collisions in our model, the Hamiltonian is not 
bilinear and the equations of motion for the operators of interest are no longer 
linear.  Specifically, using equation (\ref{hem}) we obtain the following 
set of differential equations
\bea
\fl\frac{d(\ak a)}{dt}=-2i\int_0^\infty dk g(k)(\ak b_k-\bk_k a),
\label{emnla}\\
\fl\frac{d(\ak b_k)}{dt}=-i(\omega_k-\omega_0)\ak b_k-ig_k\ak a+
	2i\int_0^\infty dq g(q)\bk_q b_k- 
	2i\int_0^\infty dq {\cal \kappa}(k,q)\ak b_q \ak a \nonumber\\ 
\lo     +4i\int_0^\infty dq\int_0^\infty d\pr{q} {\cal \kappa}(q,\pr{q})\bk_q b_{\pr{q}}\ak b_k,
\label{emnlb}\\
\fl\frac{d(\bk_k b_q)}{dt}=-i(\omega_q-\omega_k)\bk_k b_q + 
	i g(k)\ak b_q - i g(q)a \bk_k
	-2i\int_0^\infty d\pr{k} {\cal \kappa}(q,\pr{k})\ak a\bk_k b_{\pr{k}}\nonumber\\
\lo	+2i\int_0^\infty d\pr{k} {\cal \kappa}(\pr{k},k)\ak a\bk_{\pr{k}} b_q,
\label{emnlc}
\eea
which can not be solved by means of the Laplace transform. 

In order to proceed to the solution of the above equations in the context of the 
discretization approach, we have to incorporate appropriately the interatomic 
collisions in the discretization schemes developed earlier in section \ref{secII}. 
To this end, we assume that the interatomic coupling spans the same discretized 
space with the output coupling and thus does not affect the far off-resonant 
$b_k$ operators. This assumption, on the one hand, allows us  to adiabatically 
eliminate the same $b_k$ as before, while on the other to use the same 
discrete modes for both the output and the interatomic couplings. 
Note that in the case of trap-trap interatomic interactions, an adiabatic
elimination procedure of the high-lying modes has been shown to lead to the
renormalization of the (bare) interatomic potential to an effective one given
by the usual pseudopotential of equation (29) \cite{Prouk,Morgan}. We believe it is 
therefore consistent here to adiabatically eliminate high-lying modes 
(of the output spectrum) to obtain s-wave scattering between trapped and outcoupled 
atoms.
Consideration of low-energy $s$-wave scattering for interatomic 
collisions implies that our discretization should be applied under the constrain 
$k_{up}a_s<<1$.  
To confirm this, note that for the parameters used in the paper, 
$k_{up}\approx 10^6m^{-1}$ and $a_s\approx 10^{-9}m$ and thus 
$k_{up}a_s\approx 10^{-3}<<1$.

Extending equations (\ref{disc1c}) and (\ref{disc2a}) to the two-dimensional $(k_m,k_j)$ 
space we introduce the 
following coupling constants which involve pairs of modes:
\be
{\cal G}_{mj}^2={\cal M}_{G}(\omega_m,\omega_j)\delta\omega\delta\omega.
\label{discnl1}
\ee
\be
{\cal G}_{int}^2N^2=\int_{\omega_{low}}^{\omega_{up}}d\pr{\omega}\int_{\omega_{low}}^{\omega_{up}}d\omega {\cal M}_{BB}(\omega,\pr{\omega}),
\label{discnl2}
\ee
and correspond to the uniform and the non-uniform discretization scheme, respectively.
We thus obtain the following set of equations of motion for the expectation 
values of the operators involving the discrete part of the continuum
\bea
\fl \frac{d\av{\ak a}}{dt}=2\sum_{j=1}^N {\cal G}_j\Im\{\av{\ak b_j}\},
\label{emnld}\\
\fl \frac{d\av{\ak b_j}}{dt}=-i(\omega_j-\omega_0+S)\av{\ak b_j}-
	i{\cal G}_j\av{\ak a}+
	i\sum_{m=1}^N {\cal G}_m\av{\bk_m b_j}
	-i\sum_{m=1}^N {\cal G}_{jm}\av{\ak b_m \ak a}\nonumber \\
\lo	+i\sum_{l,m=1}^N {\cal G}_{lm}\av{\bk_l b_m \ak b_j},
\label{emnle}\\
\fl \frac{d\av{\bk_m b_j}}{dt}=-i(\omega_j-\omega_m)\av{\bk_m b_j} + 
	i {\cal G}_m\av{\ak b_j} - i {\cal G}_j\av{\ak b_m}^*
	-i\sum_{l=1}^N {\cal G}_{jl}\av{\ak a\bk_m b_l}\nonumber \\
\lo	+i\sum_{l=1}^N {\cal G}_{lm}\av{\ak a\bk_l b_j},
\label{emnlf}
\eea
which is not closed. 
The structure of these equations is a typical case of equations of motion emerging from 
Hamiltonians involving terms of order higher than bilinear. The main feature is the 
presence in the right hand side of operators of order higher than the one whose 
derivative is considered. Consideration of differential equations for those higher 
order correlation functions leads to the appearance of terms of even higher order 
and so on. The system of equations, in other words, does not close and there are no 
general exact remedies. 

One way to obtain an approximate solution is to decorrelate the higher order 
correlation functions into products of lower ones. Again, there usually is 
more than one way of doing so, the only guide being the demand for a closed system 
of equations. Perhaps the most straightforward decorrelation suggesting itself in 
the system of equations (\ref{emnld})-(\ref{emnlf}) is to decorrelate products of four 
operators in a way such that the 
decorrelated parts involve equal number of raising and lowering operators i.e.,
$\av{\ak b_m \ak a}\approx\av{\ak a}\av{\ak b_m}$,
$\av{\bk_l b_m \ak b_j}\approx\av{\bk_l b_m}\av{\ak b_j}$ and 
$\av{\ak a\bk_m b_l}\approx\av{\ak a}\av{\bk_m b_l}$.
We thus have the set of equations A:
\bea
\fl\frac{d\av{\ak a}}{dt}=2\sum_{j=1}^N {\cal G}_j\Im\{\av{\ak b_j}\},
\label{emnl1a}\\
\fl\frac{d\av{\ak b_j}}{dt}=-i(\omega_j-\omega_0+S)\av{\ak b_j}-
	i{\cal G}_j\av{\ak a}+
	i\sum_{m=1}^N {\cal G}_m\av{\bk_m b_j}
	-i\av{\ak a}\sum_{m=1}^N {\cal G}_{jm}\av{\ak b_m}\nonumber \\
\lo	+i\av{\ak b_j}\sum_{l,m=1}^N {\cal G}_{lm}\av{\bk_l b_m},
\label{emnl1b}\\
\fl\frac{d\av{\bk_m b_j}}{dt}=-i(\omega_j-\omega_m)\av{\bk_m b_j} + 
	i {\cal G}_m\av{\ak b_j} - i {\cal G}_j\av{\ak b_m}^*
	-i\av{\ak a}\sum_{l=1}^N {\cal G}_{jl}\av{\bk_m b_l}\nonumber \\
\lo	+i\av{\ak a}\sum_{l=1}^N {\cal G}_{lm}\av{\bk_l b_j}.
\label{emnl1c}
\eea
Applying the commutation relation $[b_m,\bk_n]=\delta_{mn}$ in 
equation (\ref{emnlf}) and decorrelating in the same fashion 
the resulting equation, we have an alternative form for equation (\ref{emnl1c}): 
\bea
\fl\frac{d\av{\bk_m b_j}}{dt}=-i(\omega_j-\omega_m)\av{\bk_m b_j} + 
	i {\cal G}_m\av{\ak b_j} - i {\cal G}_j\av{\ak b_m}^*
	-i\av{\ak b_m}^*\sum_{l=1}^N {\cal G}_{jl}\av{\ak b_l}\nonumber \\
\lo	+i\av{\ak b_j}\sum_{l=1}^N {\cal G}_{lm}\av{\ak b_l}^*,
\label{emnl2c}
\eea
which combined with equations (\ref{emnl1a}) and (\ref{emnl1b}) constitute another 
closed set of differential equations (set B).
 
\subsection{Dynamics}
\begin{figure}
  \begin{center}
    \leavevmode
    \epsfxsize8.5cm
    \epsfbox{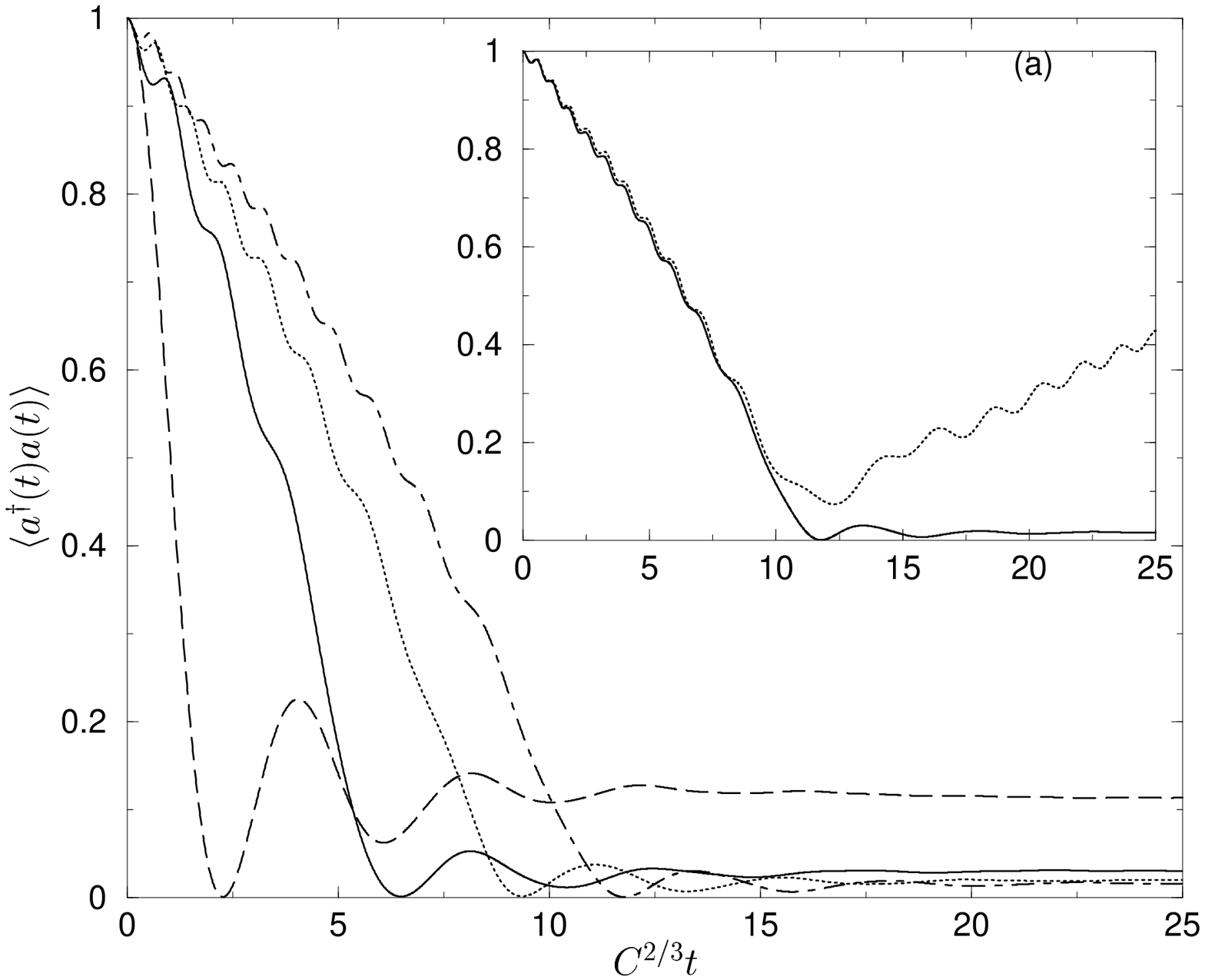}
  \end{center}
  \begin{center}
    \leavevmode
    \epsfxsize8.5cm
    \epsfbox{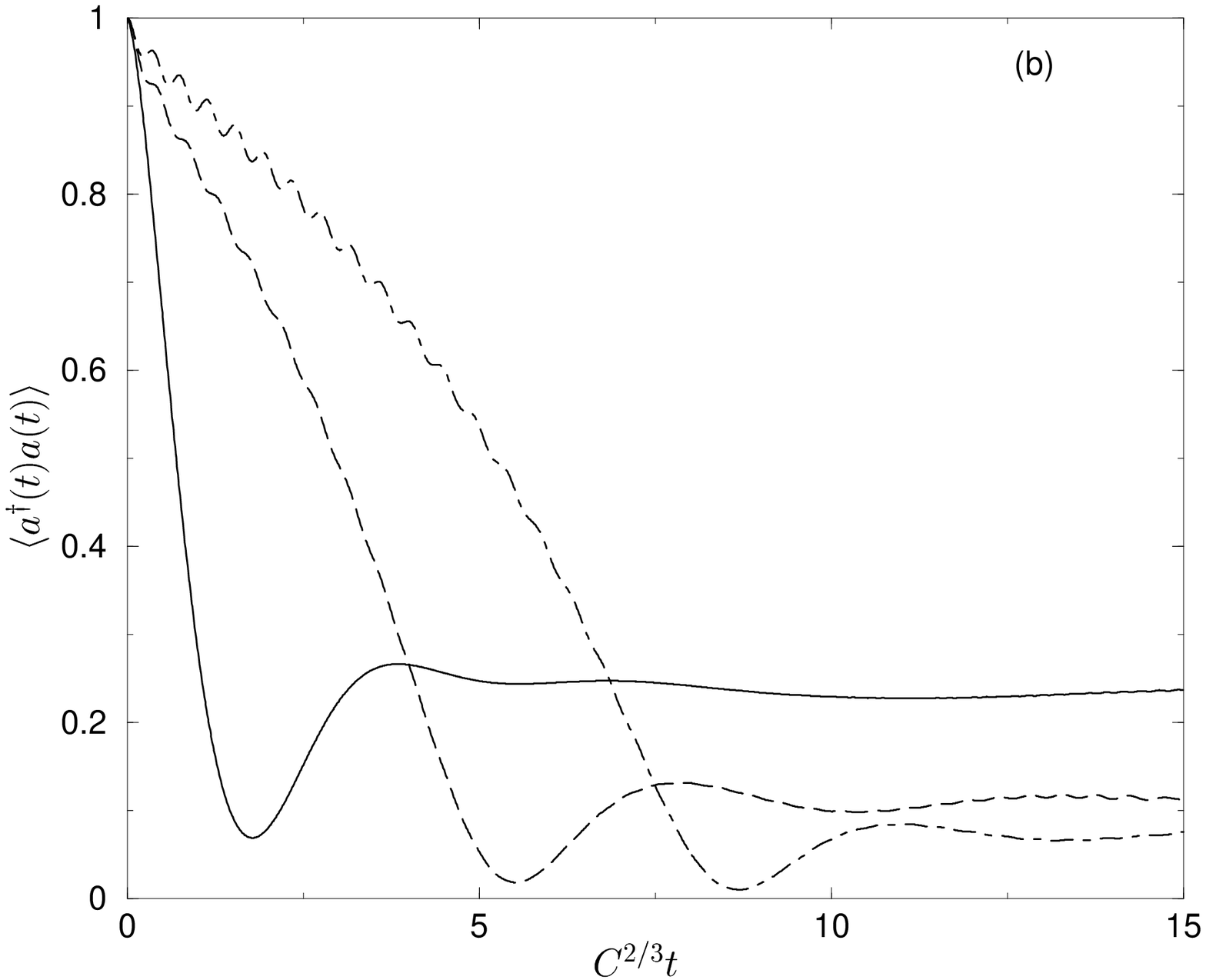}
  \end{center}
\caption{
	Effect of trap-free interatomic interactions on the time-evolution 
	of the normalized trap population  
	for Gaussian (a) and broad-band (b) couplings. 	
	All calculations have been obtained in the context of set B.
	(a) The coupling constant is $\Gamma=10^6 sec^{-2}$ and the 
	population is plotted for various interatomic-coupling strengths:
	$C_M=0$ (long-dashed line), $C_M=4C^{2/3}$ (solid line), 
	$C_M=8C^{2/3}$ (dotted line) and  
	$C_M=12C^{2/3}$ (dot-dashed line).
	Discretization parameters: $N=60$, $\omega_{low}=0$,   
	$\omega_{up}=3C^{2/3}$.
 	Inset: The solid and the dotted lines correspond to a propagation of 
	set B and equations (\ref{emdiag1})-(\ref{emdiag3}) respectively, 
	for $\Gamma=10^6 sec^{-2}$ and $C_M=12C^{2/3}$.
	(b) The solid line is for $C_M=0$, the 
	dashed line for $C_M=1C^{2/3}$ and the dot-dashed line for 
	$C_M=2C^{2/3}$. 
	Discretization parameters: $N=60$, $\omega_{low}=0$, 
	$\delta\omega=0.004C^{2/3}$. 
}
\label{variusCM_G.fig}
\end{figure}

In the absence of interatomic collisions the effective coupling constant $C$ is 
the one that determines the time scale of the initial transient regime i.e., 
the speed at which the trap-mode loses population for short time. In the present 
case, however, interatomic collisions are expected to be the dominant processes 
for short times, determining thus the corresponding time scale.  
Both sets of equations (A and B) 
predict the same dynamics for the system. In figure \ref{variusCM_G.fig}, 
we plot the solutions for the population in the 
trap as a function of the dimensionless time $C^{2/3}t$ obtained through the set 
of equations B. Clearly, the rate at which 
the trap loses population decreases with increasing values of $C_M$ 
while the trap-mode excitation undergoes oscillations which reflect the interference 
between the two possible decay routes for the system, corresponding to 
$\omega_0$ and $\omega=0$, respectively. 

\begin{figure}
  \begin{center}
    \leavevmode
    \epsfxsize8.5cm
    \epsfbox{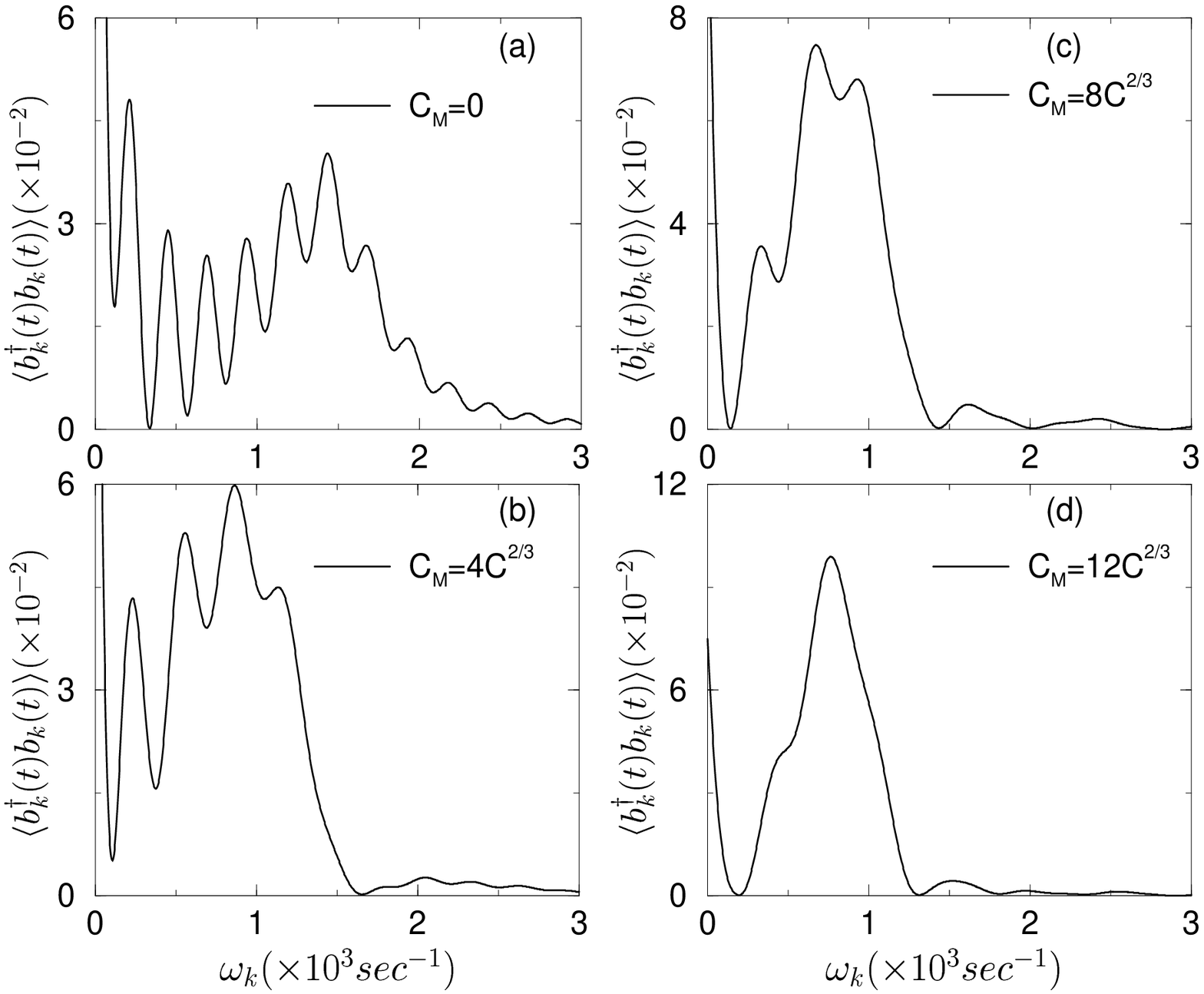}
  \end{center}
  \begin{center}
    \leavevmode
    \epsfxsize8.5cm
    \epsfbox{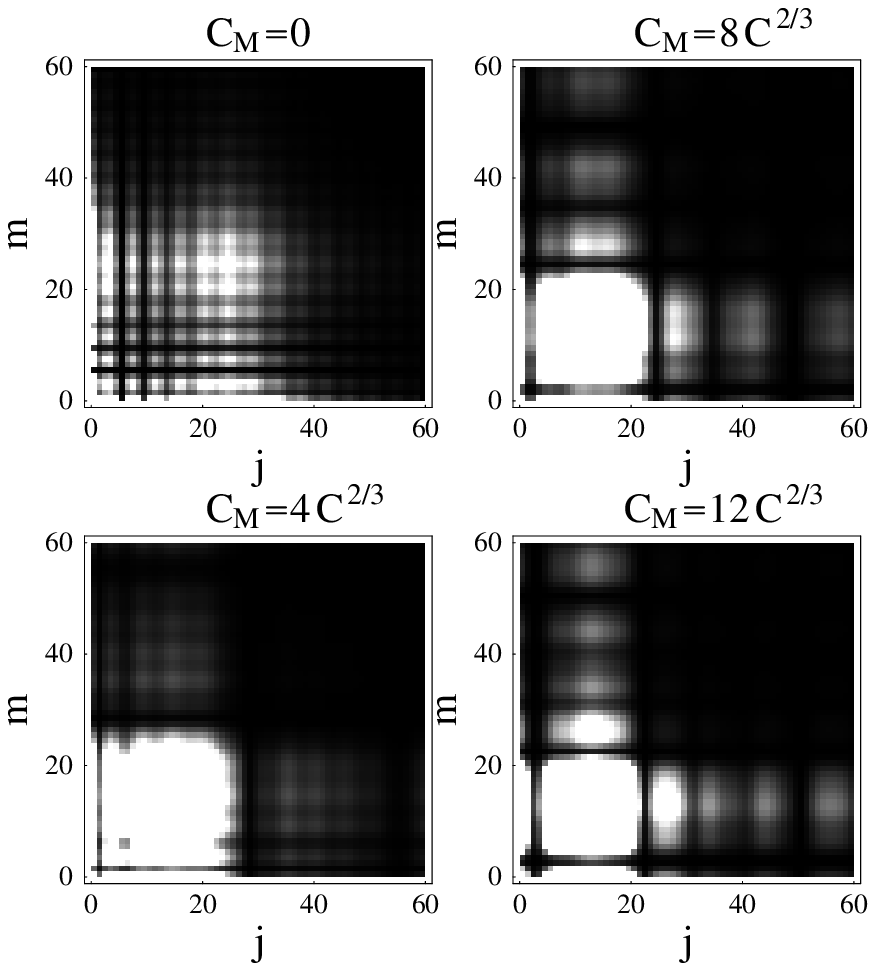}
  \end{center}
\caption{
	Effect of trap-free interatomic interactions on the energy distribution of 
	the outcoupled atoms (upper graphs) and 
	the magnitude of the matrix elements $\av{\bk_m(t)b_j(t)}$ (lower graphs). 
	All the graphs are for Gaussian coupling $(\Gamma=10^6 sec^{-2})$ and 
	correspond to $C^{2/3}t=25$ and various interatomic-coupling strengths.
	Dark and white regions in the lower graphs represent 
	matrix elements with negligible and large magnitude respectively. 
	Discretization parameters as in figure \ref{variusCM_G.fig}. 
}
\label{coherenceG2.fig}
\end{figure}

The functional dependence of ${\cal M}(\omega,\pr{\omega})$ on $\omega$ and  
$\pr{\omega}$, respectively, reveals the main role of interatomic collisions, namely 
the decoherence. We know already (section \ref{secII}) that the coherence between the 
atoms is associated with the non-Markovian behavior of the system and the 
partial depletion of the excitation of the initially highly populated 
condensate-mode. Under the influence of interatomic collisions, however, the 
coherence between the outcoupled atoms is reduced and is restricted to the vicinity of 
the trap-mode frequency (see lower graphs in figure \ref{coherenceG2.fig}), 
leading thus to a significant decay of the initially trapped atomic population, 
as well as a narrow distribution of the energies of the outcoupled atoms 
around $\omega_0$ [figures \ref{coherenceG2.fig}(a)-(d)].
This is in analogy to the behavior of the system in the absence of interatomic 
collisions and for weak output couplings 
(see lower graphs in figure \ref{coherenceG.fig}, for $\Gamma=10^4-10^5 sec^{-2}$). 
It could thus be argued that the dynamics of the system are governed by 
a set of equations of motion of the form 
\bea
\fl\frac{d\av{\ak a}}{dt}=2\sum_{j=1}^N {\cal G}_j\Im\{\av{\ak b_j}\},
\label{emdiag1}\\
\fl\frac{d\av{\ak b_j}}{dt}=-i(\omega_j-\omega_0+S)\av{\ak b_j}-
	i{\cal G}_j\av{\ak a}+
	i {\cal G}_j\av{\bk_j b_j}
	-i\av{\ak a}\sum_{m=1}^N{\cal G}_{jm}\av{\ak b_m}\nonumber \\
\lo	+i\av{\ak b_j}\sum_{m=1}^N {\cal G}_{mm}\av{\bk_m b_m},
\label{emdiag2}\\
\fl\frac{d\av{\bk_m b_m}}{dt}= 
	-2{\cal G}_m\Im\{\av{\ak b_m}\}
	-2\sum_{l=1}^N {\cal G}_{lm}\Im\{\av{\ak b_m}\av{\ak b_l}^*\},
\label{emdiag3}
\eea
where no coherences between the outcoupled atoms are involved. 
As depicted in the inset of figure \ref{variusCM_G.fig}, this argument is indeed 
true for short times and relatively large interatomic couplings, where interatomic 
collisions dominate over the output coupling. 
As was expected, however, the above set of 
equations fail to describe the decoherence in our model for larger times. 
Alternative decorrelation schemes and their predictions about the system dynamics 
are discussed in the Appendix.

\newpage

\section{SUMMARY} 
We have presented an alternative approach to a simple model for atom laser
outcoupling which is capable of circumventing certain mathematical difficulties
arising from the breakdown of the Born and Markov approximations,
inherent in essentially any model of the process. The approach is computational,
relying on the discretization of the continuum representing the
reservoir of the outcoupled atoms and has been recently developed in the context of PBG
continua.
To provide a calibration of the possibilities offered by the method, we presented
representative results, some of which serve as a basis for comparison with previous
work relying on different methods, while others demonstrate how this approach can 
go beyond the above mentioned limitations.

In particular, we have studied the dynamics of a single-mode trapped condensate which,
in order to form an atom laser, is outcoupled by both Gaussian and 
broad-band couplings. In the model considered here, there is no 
mechanism to remove the outcoupled atoms, and the ensuing coupling between trapped and 
outcoupled atoms provides (in certain limits) 
a striking manifestation of the non-markovian nature of the 
outcoupling. For example, 
considering initially a situation in which condensate and outcoupled atoms 
do not interact, we have shown that, after an initial 
transient regime where part of the trapped atomic population is lost, the system 
reaches a steady state for which the population 
inside the trap has not been totally depleted. This non-markovian nature
is most pronounced when outcoupling at a rate larger than (or comparable to) the trap 
frequency, and  is also evident in the coherences between 
trapped-outcoupled, or outcoupled-outcoupled atoms, which additionally experience a 
peak around $\omega \approx 0$.

These coherences thus play a significant role in the time evolution of the 
system. 
Perhaps more significant is the effect that the addition of interactions 
between condensate and outcoupled atoms has on the outcoupled spectrum.
It has been previously argued that interatomic interactions could 
potentially destroy 
this `bound mode' of atoms within the trap. In this work we have explicitly considered 
the effect of interactions between trapped and outcoupled atoms and have shown directly 
that realistic interaction strengths will largely destroy this `bound mode'. 
Interatomic interactions also have a drastic effects on the coherences between 
trapped and outcoupled atoms. We have thus shown that our computational approach 
is capable of exploring the entire domain of outcoupling from non-markovian to 
markovian, and can additionally include atom-atom interactions.

It should be stressed that this paper only describes the mechanism of outcoupling,
 and does not deal with the full complexity of an actual atom laser. In particular, 
we have assumed that we begin with a single-mode condensate, without any thermal 
excitations being present. Thus, strictly-speaking, our treatment is restricted to the 
zero temperature limit. 
Consideration of excitations into our model will lead firstly, to a 
decrease in the amount of coherence present in the trapped system. This
will subsequently limit the coherence transferred to the outcoupled beam
(atom laser), with the extent of the decrease from the purely
condensed case, depending on the particular choice of the outcoupling, since
some thermal component will be inevitably outcoupled. This point has been discussed 
in \cite{jap99,jap00}, whereas a three-dimensional treatment for Raman pulsed atom 
lasers has been recently performed \cite{Ruot}.

The model considered here, does not account for a mechanism of removal of the 
outcoupled atoms. Such a mechanism arises naturally from the effect of gravity, and can 
also be controlled by the amount of momentum imparted to the outcoupled atoms via Raman 
outcoupling. When the outcoupled atoms are removed from the system, the only way to 
reach the desired steady state is by continuously pumping the trapped condensate. 
Both of these features are essential in the steady-state operation of an atom laser, 
and can be included without particular difficulty in an extended version of our 
presented model. However, it bears repeating that the main aim of this paper was not to 
describe the full atom laser operation, but rather to introduce an alternative method 
for describing the process of outcoupling. A model dealing with a cw atom laser in a 
one-dimensional waveguide by means of a Langevin equation, where issues such as the 
laser linewidth can be addressed  has been discussed by one of us elsewhere 
\cite{Stochastic}.

One of the important features of the method  employed in this paper is the versatility 
in  handling any form of functional dependence of the outcoupling on the atomic 
momentum. Furthermore, it is amenable to the study of further properties of the system 
such as higher order correlation functions, albeit, as it appears for the
moment, within the limits of some form of decorrelation. Although the various 
decorrelation approximations we had to resort to have given more or less qualitatively 
compatible results, still the task of narrowing the possible choices needs to be 
addressed. One avenue in that direction would be the examination of  higher order of 
correlation functions, with the ultimate limitation being the size  of the computation, 
as the model becomes more complex. Computational possibilities, on the other hand, are 
not frozen but increase constantly. Clearly, analytically solvable models are 
indispensable in the insight they offer, but computational approaches are often 
necessary in examining details of the dynamics of a complex system.

\newpage

\appendix

\section{Alternative decorrelation schemes}
In general there is no unique way of decorrelating a system of coupled Heisenberg 
equations of motion. In the main body of the paper, we have already presented two 
different decorrelation schemes. We discuss two more in this appendix. 
Applying the commutation relation $[\bk_m,b_j]=\delta_{mj}$ in equation (\ref{emnle}), 
we obtain 
\bea
\fl \frac{d\av{\ak b_j}}{dt}=-i(\omega_j-\omega_0+S)\av{\ak b_j}-
	i{\cal G}_j\av{\ak a}+
	i\sum_{m=1}^N {\cal G}_m\av{\bk_m b_j}
	-i\sum_{m=1}^N {\cal G}_{jm}\av{\ak b_m \ak a}\nonumber \\
\lo	+i\sum_{l,m=1}^N {\cal G}_{lm}\av{b_m \bk_l \ak b_j}
	-i\av{\ak b_j}\sum_{l=1}^N {\cal G}_{ll}
\label{ap1}
\eea
which can be decorrelated as follows
\bea
\fl \frac{d\av{\ak b_j}}{dt}=-i(\omega_j-\omega_0+S)\av{\ak b_j}-
	i{\cal G}_j\av{\ak a}
	+i\sum_{m=1}^N {\cal G}_m\av{\bk_m b_j}
	-i\av{\ak a}\sum_{m=1}^N {\cal G}_{jm}\av{\ak b_m}\nonumber \\
\lo	+i\sum_{l,m=1}^N {\cal G}_{lm}\av{\ak b_m}\av{\bk_l b_j}
	-i\av{\ak b_j}\sum_{l=1}^N {\cal G}_{ll}.
\label{ap2}
\eea
This equation of motion can replace equation (\ref{emnl1b}) in the sets of equations 
we presented in the previous sections. The resulting sets of 
equations, however, predict substantially different behavior for the system 
than the one discussed up to now. Specifically, we have found that 
the number of the outcoupled atoms decreases significantly for large values of $C_M$ 
in relation to $C^{2/3}$. Thus, the atomic population of the trap tends to remain 
trapped forever $(\av{\ak(t)a(t)}\approx 1)$. Actually, this behavior stems from 
the last factor in equation (\ref{ap2}), which leads to a non-dissipative Rabi-like 
oscillation of the cavity population close to unity i.e., 
$\av{\ak(t)a(t)}\approx 1-\exp(-i\sum_l{\cal G}_{ll}t)$. 
We may overcome this problem by decorrelating the last term in equation (\ref{emnle}) as 
$\av{\bk_l b_m\ak b_j}\approx\av{\ak b_m}\av{\bk_l b_j}$ obtaining thus 
equation (\ref{ap2}) without the term involving ${\cal G}_{ll}$.    

Using the corresponding commutation relation for the cavity-mode 
operators i.e.,  $[a,\ak]=1$, we obtain from equation (\ref{emnlf})
\bea
\fl \frac{d\av{\bk_m b_j}}{dt}=-i(\omega_j-\omega_m)\av{\bk_m b_j} +
	i {\cal G}_m\av{\ak b_j} - i {\cal G}_j\av{\ak b_m}^* +
	i\sum_{l=1}^N {\cal G}_{jl}\av{\bk_m b_l} 
	-i\sum_{l=1}^N {\cal G}_{lm}\av{\bk_l b_j}\nonumber \\
\lo	-i\sum_{l=1}^N {\cal G}_{jl}\av{a\ak\bk_m b_l}+
	i\sum_{l=1}^N {\cal G}_{lm}\av{a\ak \bk_l b_j},
\label{ap3}
\eea
which after decorrelation reads
\bea
\fl \frac{d\av{\bk_m b_j}}{dt}=-i(\omega_j-\omega_m)\av{\bk_m b_j} +
	i {\cal G}_m\av{\ak b_j} - i {\cal G}_j\av{\ak b_m}^*
	+i\sum_{l=1}^N {\cal G}_{jl}\av{\bk_m b_l} 
	-i\sum_{l=1}^N {\cal G}_{lm}\av{\bk_l b_j}\nonumber \\
\lo	-i\sum_{l=1}^N {\cal G}_{jl}\av{\ak b_m}^* \av{\ak b_l}+
	i\sum_{l=1}^N {\cal G}_{lm}\av{\ak b_l}^* \av{\ak b_j}.
\label{ap4}
\eea
Although this equation of motion, accompanied by equations (\ref{emnl1a}), 
(\ref{emnl1b}), predict mainly the same short-time dynamics that we have discussed 
in section \ref{secIII}, it overestimates 
the quasi-steady state of the trap-mode as well as the oscillations in the 
cavity excitation. In other words, equation (\ref{ap4}) seems to overestimate the 
coherences in our system, due to the appearance of terms like 
$\sum_{l=1}^N {\cal G}_{lm}\av{\bk_l b_j}$ in its right-hand side.  

\ack
NPP would like to acknowledge financial support from the U.K. EPSRC.


\newpage

\Figures

\newpage

\newpage

\newpage

\newpage

\newpage

\end{document}